\documentclass[twocolumn,english,prl,floatfix,citeautoscript,nofootinbib,superscriptaddress]{revtex4}
\usepackage{amsfonts}
\usepackage{amsbsy}
\usepackage{latexsym,epsfig,graphicx}
\usepackage{dcolumn}
\usepackage{subfigure}
\usepackage{comment}
\usepackage{color}
\usepackage[colorlinks,urlcolor=blue,citecolor=blue]{hyperref}
\usepackage{amstext}
\usepackage{amssymb}
\usepackage{setspace}

\begin{document}

\title{Photonic topological insulators induced by non-Hermitian disorders in a coupled-cavity array}
\author{Xi-Wang Luo}
\thanks{luoxw@ustc.edu.cn}
\affiliation{Department of Physics, The University of Texas at Dallas, Richardson, Texas
75080-3021, USA}
\affiliation{CAS Key Laboratory of Quantum Information, University of Science and Technology of China, Hefei 230026, China}
\author{Chuanwei Zhang}
\thanks{chuanwei.zhang@utdallas.edu}
\affiliation{Department of Physics, The University of Texas at Dallas, Richardson, Texas
75080-3021, USA}

\begin{abstract}
Recent studies of disorder or non-Hermiticity induced topological insulators
inject new ingredients for engineering topological matter. Here we consider
the effect of purely non-Hermitian disorders, a combination of these two
ingredients, in a 1D coupled-cavity array with disordered gain and loss.
Topological photonic states can be induced by increasing gain-loss disorder strength
with topological invariants carried by localized states in the complex bulk
spectra. The system showcases rich phase diagrams and distinct topological
states from Hermitian disorders. The non-Hermitian critical behavior is characterized by the biorthogonal
localization length of zero-energy edge modes, which diverges at the
critical transition point and establishes the bulk-edge correspondence.
Furthermore, we show that the bulk topology may be experimentally accessed
by measuring the biorthogonal chiral displacement, which can be extracted
from a proper Ramsey interferometer that works in both clean and disordered
regions. The proposed coupled-cavity photonic setup relies on techniques
that have been experimentally demonstrated,
and thus provides a feasible route towards exploring such
non-Hermitian disorder driven topological insulators.
\end{abstract}

\maketitle

\section{Introduction}
Topological insulators (TIs),
exotic states of matter exhibiting gapless edge modes determined by
quantized features of their bulk \cite%
{xiao2010berry,hasan2010colloquium,qi2011topological,RevModPhys.88.035005},
have been widely studied in various systems~\cite%
{PhysRevLett.95.146802,Bernevig2006quantum,Konig2007quantum,PhysRevLett.111.185301,PhysRevLett.111.185302, Jotzu2014Experimental, Goldman2016topological,Cooper2018topological, haldane2008possible, hafezi2011robust,fang2012realizing,lu2014topological,kraus2012topological, Hafezi2013imaging,Ozawa2018Topological, Ma2019Topological}%
. Recently, the concept of TIs has been generalized to open quantum systems
characterized by non-Hermitian Hamiltonians~\cite{Rept.Prog.Phys.70.947},
which may exhibit unique properties without Hermitian counterparts~\cite%
{Eur.Phys.J.Spec.Top227}. The experimental advances in controlling gain and
loss in photonic systems ~\cite%
{PhysRevLett.100.103904, PhysRevLett.115.200402,
PhysRevLett.115.040402,NatMater16433,Photon.Rev.6.A51,
Nature488.167,PhysRevLett.113.053604,Science346Loss,Zhao2018topological,
PhysRevLett.120.113901,Bandres2018Topological,St-Jean2017Lasing},
as well as other systems such as atomic and electric circuit systems \cite{Muller2012Engineered,Ashida2017Parity,PhysRevLett.121.026403,arXiv1802.00443,PhysRevB.98.035141,PhysRevLett.118.045701,arXiv1608.05061,
arXiv1811.06046,arXiv1907.11562}
provide powerful tools for studying non-Hermitian topological phases. Beside
new %non-Hermitian
topological invariants~\cite%
{PhysRevB.84.205128,PhysRevA.93.062101,PhysRevLett.121.026808,
PhysRevLett.116.133903, PhysRevLett.121.136802, PhysRevLett.121.213902,
PhysRevLett.121.086803,PhysRevX.8.031079}, the unique features (e.g.,
complex eigenvalues, eigenstate biorthonormality, exceptional points, etc.)
of non-Hermitian systems can lead to novel topological phenomena, such as
the non-Hermitian skin effects, exceptional rings and bulk Fermi arcs, with
bulk-edge correspondence very different from the Hermitian systems~\cite%
{PhysRevLett.116.133903, PhysRevLett.121.136802, PhysRevLett.121.213902,
PhysRevLett.121.086803,PhysRevX.8.031079,PhysRevLett.121.026808,arXiv1808.09541, PhysRevB.97.045106, PhysRevLett.118.040401,PhysRevLett.120.146402,Xiong2018why, Science359Observation, arXiv1804.04676,PhysRevB.99.081103,arXiv1812.02011, arXiv1809.02125,arXiv1902.07217, PhysRevB.100.075403,arXiv1905.07109, PhysRevLett.123.170401,PhysRevLett.123.246801,PhysRevB.100.045141,arXiv1912.04024,arXiv1910.03229,PhysRevB.97.075128,arXiv1806.10268,ncommss41467,arXiv1907.12566}%
.

A key property of TIs (either Hermitian or non-Hermitian) is their
robustness against weak disorders through the topological protection~\cite%
{xiao2010berry,hasan2010colloquium,qi2011topological}. For sufficiently
strong disorders, the system becomes topologically trivial through Anderson
localization~\cite{PhysRev.109.1492}, accompanied by the unwinding of the
bulk topology~\cite%
{PhysRevB.50.3799,PhysRevLett.98.076802,PhysRevLett.97.036808,PhysRevLett.105.115501}%
. In this context, the prediction of the reverse process that non-trivial
topology can be induced, rather than inhibited, by the addition of disorder
to a trivial insulator was surprising~\cite{PhysRevLett.102.136806}. The
disorder-induced topological states, known as topological Anderson
insulators (TAIs), can support robust topological invariants carried
entirely by localized states. They have attracted many theoretical studies~%
\cite{PhysRevB.80.165316,PhysRevLett.103.196805,PhysRevLett.105.216601,
PhysRevLett.112.206602, PhysRevLett.114.056801,PhysRevLett.119.183901,
PhysRevLett.113.046802,PhysRevB.89.224203} and been experimentally
demonstrated in 1D synthetic atomic wires~\cite{science.aat3406} and 2D
photonic lattices~\cite{Nature560.461}.

So far the studies of TAIs have been mainly focused on Hermitian disorders,
where a major challenge for their implementation is
the average of
large numbers of disorder configurations due to the non-tunability of the
fabricated devices.
In contrast, non-Hermitian disorders through gain and loss can be tuned
through additional pumping in photonic
system and different disorder configurations can be realized
on a single optical device.
Therefore, two natural questions arise: Can novel topological
states be induced by purely non-Hermitian disorders? If so, what are the
criticality and bulk-edge correspondence for general non-Hermitian TAIs?
Furthermore,
experimental schemes for probing non-Hermitian bulk topology are highly
desirable, but still lacked even for clean systems.

In this paper, we address these important issues by considering a 1D chiral
symmetric lattice in the presence of purely non-Hermitian disorders (e.g.,
disordered gain and loss), and develop feasible experimental implementations
based on photons in coupled cavity arrays. Our main results are:
(i) Photonic topological insulators can be induced
solely by gain-loss disorders with topological winding number carried
by localized states in the (complex) bulk spectra. Such non-Hermitian TAIs
reveal richer phase diagrams and distinct topological states compared to the
Hermitian one. (ii) We examine the general critical behaviors of
non-Hermitian TAIs by deriving the biorthogonal localization length of the
zero edge mode, which diverges at the critical transition point,
establishing the bulk-edge correspondence. (iii) We propose to
experimentally probe the bulk topology by measuring the biorthogonal chiral
displacement $\mathcal{C}$, which converges to the winding number %$\nu$
under time-averaging [in the parity-time (PT) symmetric region]. We show how
$\mathcal{C}$ can be extracted using a proper Ramsey interferometer, which
provides the first realistic method to measure the topological invariants of
non-Hermitian systems and works for both disorder and clean regions.

\section{Model}
\begin{figure}[tb]
\includegraphics[width=1.0\linewidth]{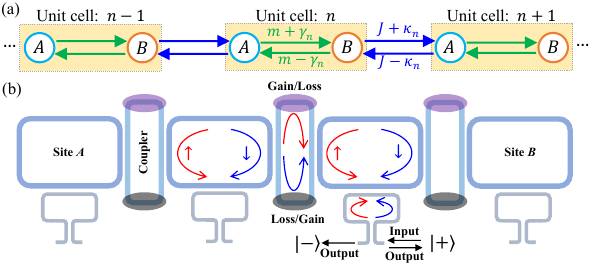}
\caption{(a) Lattice representation of the generalized SSH Hamiltonian in Eq.~%
\protect\ref{eq:H}, with purely non-Hermitian disorders $\protect\gamma_n$, $%
\protect\kappa_n$. (b) Experimental implementation of the lattice model shown in (a) using coupled arrays of cavities.}
\label{fig:lattice}
\end{figure}

We consider a disordered non-Hermitian
Su-Schrieffer-Heeger (SSH)~\cite{PhysRevLett.42.1698} model with a chiral or
sublattice symmetry, as shown in Fig.~\ref{fig:lattice}a. The tight-binding
Hamiltonian reads
\begin{eqnarray}
H&=&\sum_{n} \big[c_{n}^\dag (m_n\sigma^x-i\gamma_n\sigma^y) c_n  \nonumber
\\
& &+J^+_n c_{n}^\dag \sigma^+ c_{n+1}+J^-_n c_{n+1}^\dag \sigma^- c_{n}\big]
\label{eq:H}
\end{eqnarray}
with $c_n^\dag=(c_{n,A}^\dag,c_{n,B}^\dag)$ the particle creation operator
of sites $A$ and $B$ in the unit cell $n$ ($n\in[-N,N-1]$), and $\sigma^\pm=%
\frac{1}{2}(\sigma^x\pm i\sigma^y)$, $J^\pm_n=J_n\pm \kappa_n$. $m_n$ and $%
J_n$ are the Hermitian parts of the intra- and inter-cell tunnelings, which
are set to be uniform $m_n=m$ and $J_n=J$. The purely non-Hermitian
disorders are given by the anti-Hermitian parts $\gamma_n$ and $\kappa_n$
(corresponding to gain and loss during tunnelings), which are independently
and randomly generated numbers drawn from the uniform distributions $[\gamma_%
\text{b}-\frac{W_\gamma}{2},\gamma_\text{b}+\frac{W_\gamma}{2}]$ and $%
[\kappa_\text{b}-\frac{W_\kappa}{2},\kappa_\text{b}+\frac{W_\kappa}{2}]$,
respectively. The biases $\gamma_\text{b}$ and $\kappa_\text{b}$ correspond
to the periodic non-Hermiticities. The model preserves the chiral symmetry $%
\Gamma H \Gamma^{-1}=-H$ ($\Gamma$ flips the sign of particles on sites $B$%
), and thus the eigenvalues appear in pairs $(E,-E)$.
The system also preserves a hidden PT symmetry (see Appendix A), which may be
broken by strong non-Hermitian disorders. Hereafter, we will set $J=1$ as
the energy unit and focus on $\gamma_\text{b}=0$ unless otherwise noted.

The above disordered
non-Hermitian SSH model can be realized using photons in
coupled optical cavities~\cite%
{PhysRevLett.120.113901,St-Jean2017Lasing,Bandres2018Topological,Zhao2018topological,Sci.Rep.5.13376,science.aba8996}%
.
Shown in Fig.~\ref{fig:lattice}b is the proposed photonic circuit, where
site micro-ring cavities are evanescently coupled to their nearest neighbors using a
set of auxiliary micro-rings, each of which can be controlled independently.
The non-Hermitian (asymmetric) tunnelings can be realized by adding gain and
loss to the two arms of the auxiliary micro-rings, respectively~\cite%
{Bandres2018Topological,Zhao2018topological,Sci.Rep.5.13376,science.aba8996}%
.
We first consider the clockwise modes only, for either leftward or rightward tunnelings, photons
in the micro-ring cavities use different arms of the
auxiliary micro-rings with opposite non-Hermiticities, leading to exactly the Hamiltonian $H$
in Eq.~\ref{eq:H} (see Appendix B for more details). Similarly, the counterclockwise modes
are characterized by Hamiltonian $H^{\dag}$ which will be used to probe the bulk topology,
as we discussed later.

\section{Results}
\textbf{{Phase diagrams}.}
For disordered systems, the
topological invariant should be defined in real space. Even in the clean
limit, the real-space topological invariant would be a natural choice to
characterize non-Hermitian systems %This is because,
due to their extreme sensitivity to boundary conditions (open-boundary
spectra are very different from the periodic ones), and the topology
is encoded in the spectra with open-boundary condition where protected edge
states can exist.

We generalize the real-space winding number~\cite{PhysRevLett.113.046802}
for a non-Hermitian chiral-symmetric system as (see Appendix C)
\begin{eqnarray}
\nu&=&\frac{1}{4}\mathcal{T}\{Q\Gamma[\hat{X},Q]+Q^\dag\Gamma[\hat{X},Q^\dag]%
\},  \label{eq:winding}
\end{eqnarray}
where $\hat{X}$ is the unit-cell position operator, $\mathcal{T}$ is the
trace per volume and $Q=P_+-P_-$ is the flattened Hamiltonian with $%
P_\pm=\pm\sum_{j}|\Psi^R_{j,\pm}\rangle\langle\Psi^L_{j,\pm}|$. $%
|\Psi^{R,L}_{j,\pm}\rangle$ satisfy $H|\Psi^R_{j,\pm}\rangle=E_{j,\pm}|%
\Psi^R_{j,\pm}\rangle$ and $H^\dag|\Psi^L_{j,\pm}\rangle=E^*_{j,\pm}|%
\Psi^L_{j,\pm}\rangle$, with chiral-symmetric pairs $E_{j,+}=-E_{j,-}$ and $%
|\Psi^{R,L}_{j,\mp}\rangle=\Gamma|\Psi^{R,L}_{j,\pm}\rangle$. The choice of
occupied bands $E_{j,-}$ is very flexible and different choices lead to the
same winding number. Different from Hermitian TAIs, here both $Q$ and $%
Q^\dag $ ($Q\neq Q^\dag$) are included in the definition of the winding
number to guarantee a real-valued invariant.

\begin{figure}[tb]
\includegraphics[width=1.0\linewidth]{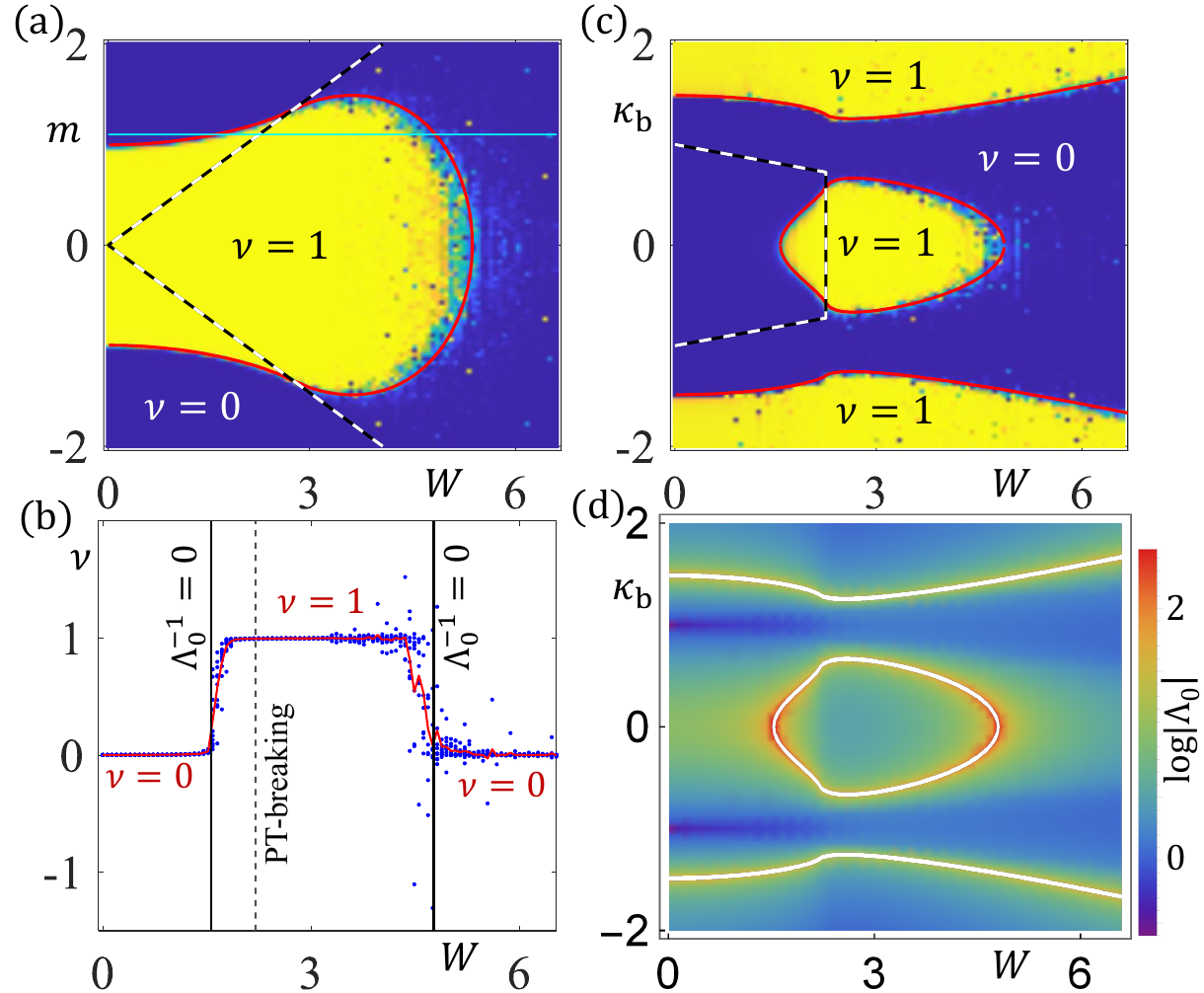}
\caption{(a) Phase diagram in the $m$-$W$ plane with $\protect\kappa_\text{b}%
=0.1$. The winding number (shown by color scale) is
averaged over 10 disorder configurations for an extremely large $N=300$.
Red solid line is the
analytic critical boundary and stripped line is the PT-symmetry breaking
curve. (b) The winding number along the horizontal line in (a) with $m=1.1$.
The un-averaged data for 10 disorder configurations are shown by the
scattered blue points and the average by the red solid line. (c) Phase
diagram in the $\protect\kappa_\text{b}$-$W$ plane with $m=1.1$. Other
parameters are the same as in (a). (d) The analytic zero-mode
Biorthogonal localization length corresponding to the phase diagram in (c). $%
W_\protect\gamma=8W_\protect\kappa=W$ in all plots.}
\label{fig:phase}
\end{figure}

We first consider stronger intra-cell disorders $W_\gamma=8W_\kappa=W$. Fig.~%
\ref{fig:phase}a shows the phase diagram
%determined by numerical calculation of $\nu$
in the $W$-$m$ plane with $\kappa_\text{b}=0.1$. At $W=0$ (no
disorder), the system is in topological phase with $\nu=1$ for $%
m^2<|1-\kappa_\text{b}^2|$~\cite{PhysRevLett.121.086803}. For a small $W$,
the disorders tend to weaken the intra-cell couplings, leading to the
enlargement of the topological region as $W$ increases. That is, trivial
system becomes topological through the addition of non-Hermitian disorders,
as shown in Fig.~\ref{fig:phase}b with $m=1.1$.
%(corresponding to the horizontal line in Fig.~\ref{fig:phase}a).
The winding number fluctuates
strongly near the phase boundary and the fluctuations are more significant
for stronger non-Hermiticities (see Appendix C). In the $W\rightarrow\infty$ limit,
the intra-cell couplings become dominant and the system must become trivial (%
$\nu=0$) for all $m$ and $\kappa_\text{b}$.
%as indicated by the dashed line.
In Fig.~\ref{fig:phase}c, we plot the phase diagram as a function of $\kappa_%
\text{b}$ for $m=1.1$. The area of disorder-induced topological phase
shrinks to zero as $|\kappa_\text{b}|$ increases, leading to a topological
island in the $W$-$\kappa_\text{b}$ plane. In the strong $|\kappa_\text{b}|$
limit, the system enters the topological phase again as the inter-cell
couplings becomes dominant. Such phase diagrams are unique for non-Hermitian
disorders, since the interplay between Hermitian and non-Hermitian
tunnelings tends to weaken each other
%and as we tune the two-site tunneling from Hermitian
%to non-Hermitian limit,
and the effective two-site coupling reaches its minimum when they are
comparable.

As $W$ increases, the gap at $E=0$ closes prior to the phase transition to $%
\nu=0$ and $\nu=1$ for $m^2<|1-\kappa_\text{b}^2|$ and $m^2>|1-\kappa_\text{b}^2|$,
respectively (see Appendix A). The stripped lines in Figs.~\ref{fig:phase}a-c are the
%boundary
%of
PT-symmetry breaking curves, indicating the topological phase is
unaffected by the disorder-driven PT-symmetry breaking or band gap closing,
%(the gap closes prior to the phase transition),
and $\nu$ remains robust for much stronger disorders. For dominant
inter-cell disorders $W_\kappa\gg W_\gamma$, the phase diagram can be
obtained similarly, where the system becomes topological, rather than
trivial, in the $W_\kappa\rightarrow\infty$ limit.

\begin{figure}[tb]
\includegraphics[width=1.0\linewidth]{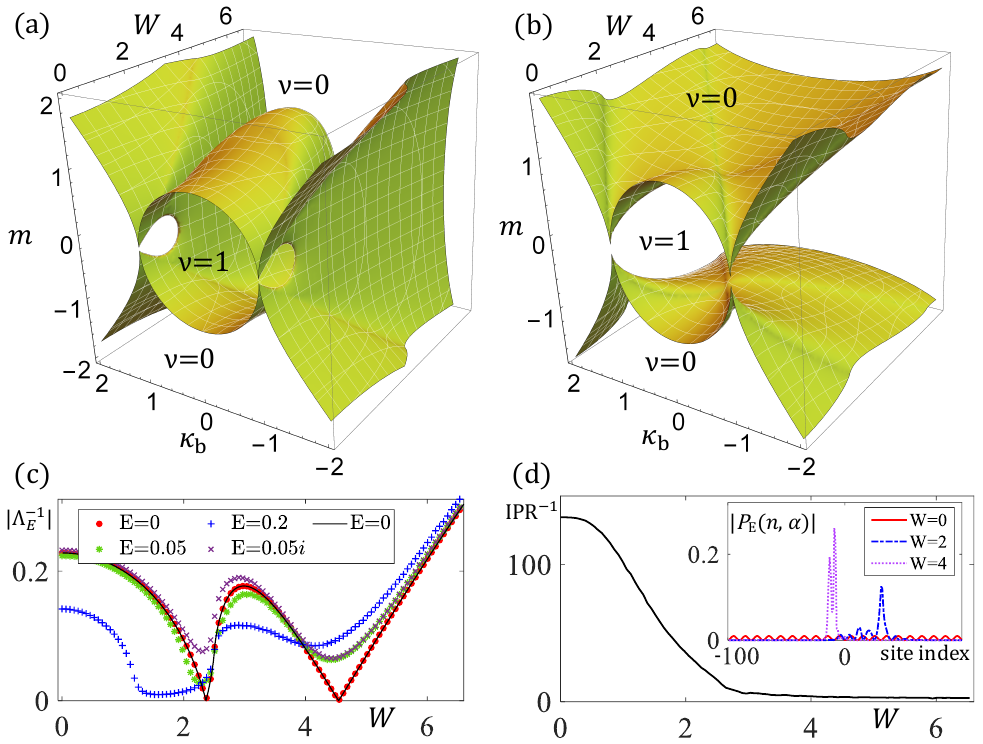}
\caption{(a) and (b) Phase diagrams in the three-dimensional parameter space
($m$, $\protect\kappa_b$, $W$) based on delocalized critical
surface with $W=W_\protect\gamma=8W_\protect\kappa$ and $W=W_\protect\kappa%
=8W_\protect\gamma$, respectively. (c) The biorthogonal localization lengths
versus disorder at different energies. Data points are obtained by numerical
transfer matrix analysis and solid line is the analytic result for $E=0$.
The transfer matrix was iterated $10^{6\sim8}$ times. (d) Biorthogonal IPR
averaged over all states and 100 disorder configurations with %for a 200-site lattice
$N=50$. The inset shows the biorthogonal density distribution $P_E(n,%
\protect\alpha)$ around $\text{Re}[E]=-1.5$. The site index is $2n+\protect%
\alpha$. Other parameters in (c) and (d): $m=1.25$, $\protect\kappa_\text{b}%
=0.1$ and $W=W_\protect\gamma=8W_\protect\kappa$.}
\label{fig:localization}
\end{figure}

%\emph{\textcolor{blue}{Bulk-edge correspondence}.---}
\textbf{{Biorthogonal criticality}.}
For both clean and disordered
Hermitian systems~\cite{PhysRevLett.113.046802},
% It is known that
the edge states (i.e., zero-energy modes) become delocalized at the
topological critical point~\cite{RevModPhys.88.035005}.
%which applies to both clean and disordered Hermitian systems~\cite{PhysRevLett.113.046802}.
%Such critical behavior also apply to the disordered Hermitian systems.
Different from Hermitian systems, here the left and right eigenstates are
inequivalent and may suffer skin effects,
both of which need be taken into account to characterize the
criticality (see Appendix D). We examine such biorthogonal criticality by deriving
the analytic formula of the zero-mode localization length, which
enables us to identify the topological phase boundary. The zero modes
%with $E=0$
of the eigenequations $H|\Psi _{0}^{R}\rangle =0$ and $H^{\dag }|\Psi
_{0}^{L}\rangle =0$ can be solved exactly, from which we obtain the
biorthogonal distributions $P(n,\alpha )=\langle \Psi _{0}^{L}(n,\alpha
)|\Psi _{0}^{R}(n,\alpha )\rangle $ as
\begin{equation}
P(n,\alpha )=P(0,\alpha )\prod_{n^{\prime }=0}^{n-1}\left( \frac{m-\gamma
_{n^{\prime }+\alpha }^{2}}{1-\kappa _{n^{\prime }}^{2}}\right) ^{\eta
_{\alpha }}.
\end{equation}%
%for the unit cell $n$.
Here $\alpha =0,1$ and $\eta _{\alpha }=\pm $
correspond to sublattice sites $A$ and $B$, respectively. The biorthogonal
localization lengths are defined as $\Lambda _{\alpha }^{-1}=-\frac{1}{2}%
\lim_{n\rightarrow \infty }\frac{1}{n}\ln |\frac{P(n,\alpha )}{P(0,\alpha )}%
| $, which do not suffer skin effects and satisfy $\Lambda
_{0}^{-1}=-\Lambda _{1}^{-1}$. %\begin{equation}
$\Lambda _{0}$ as a function of $m$, $W_{\gamma }$, $W_{\kappa }$, $\gamma _{%
\text{b}}$ and $\kappa _{\text{b}}$
can be obtained after ensemble average (the explicit expression can be found
in Appendix D), and the critical exponent is 1. Interestingly, $\Lambda _{0}$
at $\gamma _{\text{b}}=\kappa _{\text{b}}=0$ is exactly the same as that of
the Hermitian system studied in~\cite{PhysRevLett.113.046802}, though the
characterization is different.

The bulk $\nu$ and edge $\Lambda_0$ quantities establish the generalized
bulk-edge correspondence for the non-Hermitian TAIs. In the topological
(trivial) phase with $\nu=1$ ($\nu=0$), we have $(-1)^\alpha\Lambda_\alpha>0$
%and $\Lambda_1^{-1}<0$
[$(-1)^\alpha\Lambda_\alpha<0$].
%($\Lambda_0^{-1}<0$ and $\Lambda_1^{-1}>0$).
Recall that the lattice starts (ends) by a site $A$ (B) at the left (right)
boundary, $(-1)^\alpha\Lambda_\alpha>0$ %and $\Lambda_1^{-1}<0$
indicates one zero mode at each boundary. The topological phase
transition occurs at the delocalized critical surface where the biorthogonal
localization length $\Lambda_0$ diverges (i.e., $\Lambda_0^{-1}$ crosses
zero). Fig.~\ref{fig:phase}d shows the corresponding $\Lambda_0$ of the
phase diagram in Fig.~\ref{fig:phase}c. In Figs.~\ref{fig:localization}a and %
\ref{fig:localization}b, the exact phase diagrams in the whole parameter
space $(m,\kappa_\text{b},W)$ are plotted for $\Lambda_0^{-1}=0$ with
dominant intra-cell ($W=W_\gamma=8W_\kappa$) and inter-cell ($%
W=W_\kappa=8W_\gamma$) disorders, respectively (for other disorder
configurations see Appendix D).
%We see that
The non-Hermitian TAIs  %can
support richer phase diagrams and topological phenomena going beyond
Hermitian TAIs. In Fig.~\ref{fig:localization}a, the topological regions at
small and large $|\kappa_\text{b}|$ are connected by two holes, whose size
is proportional to $W_\kappa$. In Fig.~\ref{fig:localization}b, the two
trivial regions at large $m$ are separated by the topological region around $%
m=0$. These results are in agreement with the phase diagrams predicted
by $\nu$ (see Figs.~\ref{fig:phase}a-c).

The critical behavior is characterized only by the zero modes, %and
all states with $E\neq0$ are localized in every instance with disorders.
Using a numerical analysis of the transfer matrix~\cite%
{Z.Phys.B.53,PhysRevB.29.3111} for both $H$ and $H^\dag$ (see Appendix D), we
calculate the biorthogonal localization lengths as functions of disorder
strength and confirm the localization behavior of the bulk
states (see Fig.~\ref%
{fig:localization}c). Fig.~\ref{fig:localization}d shows the disorder-averaged inverse
participation ratio (IPR)~\cite{RevModPhys.80.1355} obtained from the
biorthogonal density distributions (see Appendix D), which also imply the
localization of the entire bulk (larger IPR corresponds to stronger
localization). We emphasize that, the biorthogonal density distributions and
localization lengths do not suffer skin effects which may exist in the
left/right eigenstates.

\begin{figure}[tb]
\includegraphics[width=1.0\linewidth]{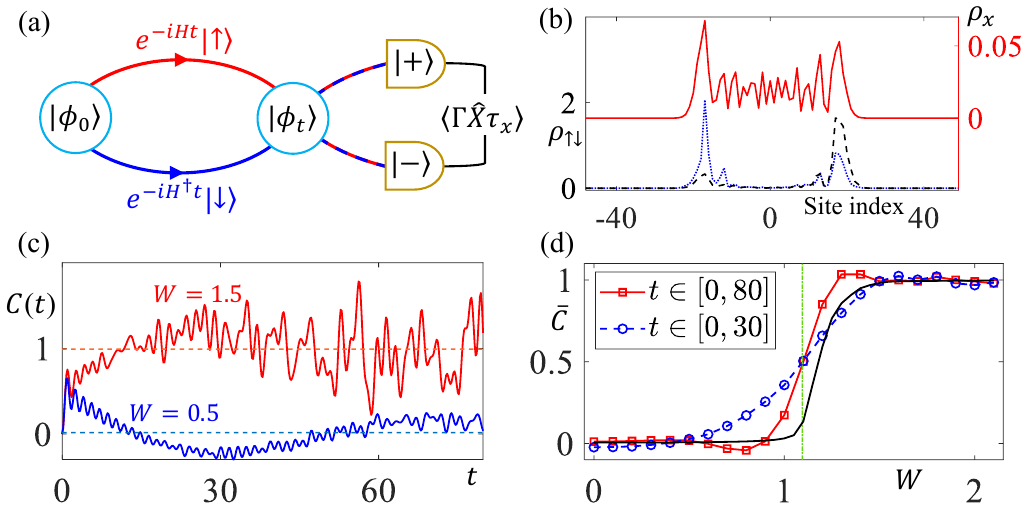}
\caption{(a) Ramsey interferometer for measuring
%biorthogonal chiral displacement
$\mathcal{C}$. (b) Density distributions of $|\protect\phi%
_t\rangle$ at $t=10$ and $W=1.5$. $\protect\rho_x$, $\protect\rho_\uparrow$
and $\protect\rho_\downarrow$ are shown by the solid, dashed and dotted
lines respectively. (c) Dynamics of $\mathcal{C}(t)$ for topological and
trivial phases at different $W$. The dashed lines are the
%corresponding
time-averaged value $\bar\mathcal{C}$ in the interval $t\in[%
0,80]$. (d) Evolution of $\bar\mathcal{C}$ with disorder
for different
averaging time intervals. The black solid line is the
corresponding winding number $\protect\nu$. All results in (b)-(d) are
averaged over 100 disorder configurations with $m=1.05$, $%
\protect\gamma_\text{b}=0$, $W=W_\protect\gamma=8W_\protect\kappa$. $N$
is chosen such that the wave functions do not reach the boundary during the
evolution.}
\label{fig:probe}
\end{figure}

\textbf{{Probing the topology}.}
For disordered system, the
zero edge modes, usually embedded in the gapless bulk spectra, are
difficult to detect; not to mention that the non-Hermitian (left or right)
bulk states may also be localized near the edges due to skin effects.
Determining the bulk winding from Eq.~\ref{eq:winding} requires the
measurement of all possible eigenstates, which is also hard to perform.
Another way to access the topological invariant is to monitor the dynamical
response of photons initially prepared on site $A$ of unit-cell $n=0$
(denoted as $|0_A\rangle$) and measure the mean chiral displacement~\cite%
{science.aat3406,ncomms15516}. Different from Hermitian models, here the
time evolution under $H$ is not unitary, and the usual scheme no longer
works. To overcome this difficulty, we introduce the biorthogonal chiral
displacement
\begin{equation}
\mathcal{C}(t)=\langle L(t)|\Gamma\hat{X}|R(t)\rangle +h.c.,
\end{equation}
which includes
the dynamics of both $H$ and $H^\dag$, with $%
|R(t)\rangle=\exp(-iHt)|0_A\rangle$, $|L(t)\rangle=\exp(-iH^\dag
t)|0_A\rangle$ and $\langle L(t)|R(t)\rangle=1$.
We find in the PT-symmetric region (i.e., real-spectrum
region), $\mathcal{C}(t)$ %the biorthogonal chiral displacement
converges to the winding number $\nu$ upon sufficient time- and
disorder-averaging $\bar\mathcal{C}=\langle\mathcal{C}(t)\rangle$ (see Appendix E). In the
clean limit, the disorder-averaging is unnecessary.

We now propose a method to measure the biorthogonal chiral displacement
using proper Ramsey interferometer sequences as shown in Fig.~\ref{fig:probe}%
a. First we introduce an
additional pseudospin degree of freedom, and prepare the photons in state $%
|\phi _{0}\rangle =|+\rangle |0_{A}\rangle $ with pseudospin state $|\pm
\rangle =\frac{1}{\sqrt{2}}(\left\vert \uparrow \right\rangle \pm \left\vert
\uparrow \right\rangle )$. Then we engineer the dissipations such that the
dynamics of photons in the two pseudospin states $\left\vert \uparrow
\right\rangle $ and $\left\vert \downarrow \right\rangle $ are governed by $H
$ and $H^{\dag }$ respectively. The total Hamiltonian reads
\begin{equation}
H_{\text{probe}}=H\otimes \left\vert \uparrow \right\rangle \left\langle
\uparrow \right\vert +H^{\dag }\otimes \left\vert \downarrow \right\rangle
\left\langle \downarrow \right\vert .  \label{eq:Hspin}
\end{equation}%
After an evolution interval $t$, we measure the chiral displacement in basis
$|\pm \rangle $, and the outcome gives
$
\mathcal{C}(t)=\langle \phi _{t}|2\Gamma \hat{X}\tau ^{x}|\phi _{t}\rangle ,
$
with $|\phi _{t}\rangle =\exp (-iH_{\text{probe}}t)|\phi _{0}\rangle $ and $%
\tau ^{x}$ the Pauli matrix in the pseudospin basis.

%As we discussed above (see Fig.~2a),
The disorder driven topological phase transition may occur before the
PT-symmetry breaking (see Fig.~\ref{fig:phase}a), and thus can be probed by
the dynamical response. %is independent from the PT-symmetry breaking.
In Fig.~\ref{fig:probe}b, we plot the typical (disorder-averaged) photonic density
distributions $\rho_s=\langle
\phi_t(n,\alpha)|\tau^s|\phi_t(n,\alpha)\rangle $
%of $|\phi_t\rangle$
in different pseudospin basis ($\tau^{\uparrow}=%
\left|\uparrow\right\rangle\left\langle\uparrow\right|$).
The pseudospin up and pseudospin down densities may be amplified with
strong asymmetric distributions away from $n=0$ %the starting site
due to the non-Hermitian (i.e., non-reciprocity) tunnelings. However, their
interference density pattern $\rho_x$, almost symmetrically distributed
around $n=0$, remains finite and normalized [$\sum_{n,\alpha}\rho_x(n,%
\alpha)=1$].
Fig.~\ref{fig:probe}c shows the disorder-averaged dynamics of $\mathcal{C}%
(t) $ with different disorder strengths, which converge to $0$ and $1$ upon
time-averaging for trivial and topological phases, respectively. The
dependence of $\bar\mathcal{C}$ on the strength of applied disorder is shown
in Fig.~\ref{fig:probe}d, which changes from $\bar\mathcal{C}=0$ to $\bar%
\mathcal{C}=1$ across the disorder driven topological phase transition.

As shown in Fig.~\ref{fig:lattice}b,
for either leftward or rightward tunnelings, the clockwise mode and
counterclockwise mode in the micro-ring cavity use different arms of the
coupler with opposite non-Hermiticities, leading to exactly the Hamiltonian $%
H_{\text{probe}}$ in Eq.~\ref{eq:Hspin} (see Appendix B), where clockwise and
counterclockwise modes play the role of pseudospin degrees of freedom. To
measure the Ramsey interference, we propose to couple each cavity with an
input-output waveguide~\cite{hafezi2011robust,PhysRevLett.120.113901}, as
shown in Fig.~\ref{fig:lattice}b. The waveguide is arranged as a Sagnac
interferometer so that the initial state $|\phi _{0}\rangle $ can be
prepared by a narrow optical input pulse applying to site $A$ at $n=0$, and the two
output ports measure the photonic pseudospin states in the basis $|+\rangle $ and $%
|-\rangle $, respectively (see Appendix B).

\section{Discussion and Conclusion}
%\emph{\textcolor{blue}{Conclusion and discussion}.---}
In summary, we
proposed and characterized non-Hermitian photonic TAIs
induced solely by gain/loss disorders which showcase richer phase diagrams and
distinct topological states compared to Hermitian TAIs, and developed the first realistic
method to probe their topological invariants.
Though we focused on the photonic systems based on coupled-cavity
arrays, it is possible to generalize our
study to other systems such as cold atoms in optical lattices and
microwaves in electric circuits, where the non-Hermitian tunnelings can be
realized by Raman couplings with lossy atomic levels~\cite%
{arXiv1608.05061,PhysRevX.8.031079,arXiv1810.04067} and circuit
amplifiers/resistances as realized in recent experiment~\cite%
{arXiv1907.11562}. Moreover, it will be exciting to study the
interaction effects and its interplay with non-Hermitian disorders, where
non-trivial many-body topological ground states or many-body localization
may exist (Recent studies suggest even periodic non-Hermiticity could
significantly alert the localization properties~\cite%
{PhysRevLett.77.570,PhysRevB.92.195107,PhysRevA.91.033815,PhysRevA.95.062118,PhysRevLett.116.237203,PhysRevLett.123.090603}%
).

Our biorthogonal %developed bulk and edge
topological characterizations may be useful for
exploring other disordered non-Hermitian systems, and
is possible to be generalized to interacting
many-body topological states (see Appendix F). The proposed Ramsey
interferometer allows the coherent extraction of information from both right
and left eigenstates, which may have potential applications in probing the
topology of various non-Hermitian systems. Previous studies
on non-Hermitian TAIs mainly focused on strong Hermitian disorders in the PT-symmetric region~\cite{arXiv1908.01172},
where topological characterization does not apply to general non-Hermitian TAIs
like our model; in addition, the critical behavior (e.g., the biorthogonal
localization length) for general non-Hermitian TAIs, the probing scheme based on chiral
displacement and its Ramsey-interferometer implementation, the coupled-cavity experimental realization were not considered.
Our work offers a new route towards exploring novel photonic topological insulators
and rich phenomena going beyond Hermitian systems and paves the way for
characterizing and probing non-Hermitian topological states.

\begin{widetext}

\section{Appendix A: Hidden PT symmetry}
\setcounter{figure}{0} \renewcommand{\thefigure}{A\arabic{figure}} %
\setcounter{equation}{0} \renewcommand{\theequation}{A\arabic{equation}}
As we mentioned in the main
text, the system obeys a hidden PT symmetry. To see this, we first apply a
unitary rotation in the sublattice space $c_n\rightarrow \widetilde{c}%
_n=\exp(i\frac{\pi}{4}\sigma^x) c_n$, and rewrite the Hamiltonian as
\begin{eqnarray}
H=\sum_{n} \big[\widetilde{c}_{n}^\dag (m\sigma^x-i\gamma_n\sigma^z)
\widetilde{c}_n +(J+\kappa_n) \widetilde{c}_{n}^\dag \frac{\sigma_x+i\sigma_z%
}{2} \widetilde{c}_{n+1}+(J-\kappa_n) \widetilde{c}_{n+1}^\dag \frac{%
\sigma_x-i\sigma_z}{2} \widetilde{c}_{n}\big].  \label{eq:H_PT}
\end{eqnarray}
Then, we define the parity and time-reversal operators in the basis $%
\widetilde{c}_n$ as $\mathcal{P}=\sigma_x$ and $\mathcal{T}=\mathcal{K}$
(with $\mathcal{K}$ denoting complex conjugation), respectively. Therefore,
the Hamiltonian preserves the PT symmetry as $\mathcal{PT}H(\mathcal{PT}%
)^{-1}=H$. The lattice representation of Eq.~\ref{eq:H_PT} is shown in Fig.~%
\ref{figS1:band}(a). The parity operator corresponds to the reflection with
respect to the horizontal dashed line, which exchanges sublattice sites $A$
and $B$. We find that the system is PT-symmetric in the weak non-Hermitian
region when both $|\gamma_n|<|m|$ and $|\kappa_n|<|J|$ are satisfied for all
$n$ (i.e., $W_\gamma+|\gamma_\text{b}|<|m|$ and $W_\kappa+|\kappa_\text{b}%
|<|J|$). Otherwise, the PT-symmetry is spontaneously broken. The spectrum is
real in the PT-symmetric region and becomes complex in the PT-symmetry
breaking region. The typical band structure is shown in Figs.~\ref%
{figS1:band}(b) and \ref{figS1:band}(c), where the spectrum becomes complex
after the PT-symmetry breaking point. As $W$ increases, the gap at $E=0$
closes prior to the phase transition to $\nu=0$ ($\nu=1$) for $m^2<|1-\kappa_%
\text{b}^2|$ ($m^2>|1-\kappa_\text{b}^2|$), and the PT-symmetry breaking is
not associated with the gap closing or topological phase transition.

\begin{figure}[tb]
\includegraphics[width=1.0\linewidth]{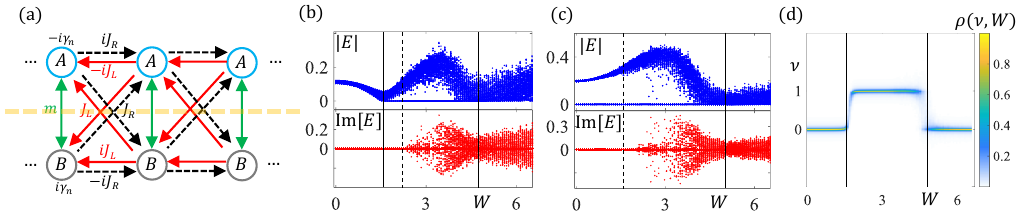}
\caption{(a) Lattice representation of the Hamiltonian in Eq.~\protect\ref%
{eq:H_PT}, with on-site loss (gain) rate $\protect\gamma _{n}$ for $A$ ($B$)
sites. The tunneling gain (loss) $\protect\kappa _{n}$ is encoded in $J_{L}=%
\frac{J-\protect\kappa _{n}}{2}$ and $J_{R}=\frac{J+\protect\kappa _{n}}{2}$%
. (b,c) The band structures around $E=0$ as a function of disorder strength
for $m=1.1$ and $m=0.8$ respectively, with $W=W_{\protect\gamma }=8W_{%
\protect\kappa }$, $\protect\kappa _{\text{b}}=0.1$. The scattered points
show the first 20 energy levels around $E=0$ with 10 disorder
configurations. The gap between the bulk and zero-energy edge states around $%
W=3$ is due to the low density of states, which will be filled if more
disorder configurations are considered. (d) The winding number distribution $%
\protect\rho (\protect\nu ,W)$ of 1000 disorder configurations. $\protect%
\rho (\protect\nu ,W)=\frac{1}{D}\sum_{d=1}^{D}\frac{\protect\varepsilon ^{2}%
}{[\protect\nu -\protect\nu _{d}(W)]^{2}+\protect\varepsilon ^{2}}$, $%
\protect\nu _{d}(W)$ is the winding number of the $d$-th disorder
configuration with disorder strength $W$, $D=1000$ is the total number of
disorder configurations, and each $\protect\nu _{d}(W)$ is replaced by a
narrow Lorentz function with width $\protect\varepsilon =0.02$. In (b-d),
the solid and dashed vertical lines are the phase boundary and PT-symmetry
breaking point, respectively.}
\label{figS1:band}
\end{figure}

\section{Appendix B: Experimental realization}
In the main text, we
have considered the realization of our model using coupled micro-ring
cavities. As we discussed in the main text, the non-Hermiticity is induced
by gain and loss in the coupler cavities. Here we give more details. Let us
consider the coupling between two site cavities. In the presence of gain and
loss in the coupler cavity, the tight-binding Hamiltonian for the clockwise
modes can be written as~\cite{Sci.Rep.5.13376}
\begin{eqnarray}
H_\text{cw}=re^{g}c_{B}^{\dag }c_{A}+re^{-g}c_{A}^{\dag }c_{B},
\end{eqnarray}%
where $r$ is determined by the coupling strength between the site and
coupler cavities, and $g$ is determined by the gain and loss strength,
leading to $m=r\cosh (g)$ and $\gamma _{n}=r\sinh (g)$. While for the
counterclockwise modes, the Hamiltonian reads
\begin{eqnarray}
H_\text{ccw}=re^{-g}c_{B}^{\dag }c_{A}+re^{g}c_{A}^{\dag }c_{B}=H_\text{cw}%
^\dag.
\end{eqnarray}%
Therefore, we obtain the total Hamiltonian of Eq.~5 in the main text, with
the clockwise and counterclockwise modes playing the role of pseudospin
degrees of freedom.

In the single coupler case, the system always stays in the PT-symmetric
region with $|\gamma _{n}|<m$. Though this does not prevent us from
observing the non-Hermitian TAIs, the system has some other shortcomings.
Once the photonic circuit is fabricated, the tunability of $m$ and $\gamma
_{n}$ is very limited. Even though gain and loss can be controlled simply by
varying the pumping strength, $m$ and $\gamma _{n}$ cannot be tuned
independently. A different disorder configuration may require additional
sample fabrication. Fortunately, all these shortcomings can be overcome by
introducing a second coupler cavity, as shown in Fig.~\ref%
{figS3:input-output}(b). By controlling the interference between two
couplers~\cite%
{Yariv.Photonics,hafezi2011robust,PhysRevLett.118.083603,PhysRevA.97.043841}%
, the Hermitian and non-Hermitian parts of the tunnelling become $m=r[\cosh
(g_{1})-\cosh (g_{2})]$ and $\gamma _{n}=r[\sinh (g_{1})-\sinh (g_{2})]$,
which can be tuned independently to arbitrary region (PT-symmetric or
PT-breaking regions) by simply gain and loss control. This allows the study
of various non-Hermitian disorder configurations using one sample.

Now we show how the input-output Sagnac waveguide allows the measurement in
the spin basis $|\pm \rangle $. As shown in Fig.~\ref{figS3:input-output}%
(a), the input-output waveguide is a ring interferometer with a $50:50$ beam
splitter (BS). Let us denote the input-output field operators as $c_{1,\text{%
in}}$, $c_{1,\text{in}}$ and $c_{1,\text{out}}$ $c_{1,\text{out}}$, their
relations with the field operators $c_{\uparrow }^{\prime }$ and $%
c_{\downarrow }^{\prime }$ inside the Sagnac interferometer are~\cite%
{Yariv.Photonics,hafezi2011robust}
\begin{eqnarray}
\left[
\begin{array}{c}
c_{\uparrow ,\text{in}}^{\prime } \\
c_{\downarrow ,\text{in}}^{\prime }%
\end{array}%
\right] =\left[
\begin{array}{cc}
\frac{1}{\sqrt{2}} & -\frac{1}{\sqrt{2}} \\
\frac{1}{\sqrt{2}} & \frac{1}{\sqrt{2}}%
\end{array}%
\right] \left[
\begin{array}{c}
c_{1,\text{in}} \\
c_{2,\text{in}}%
\end{array}%
\right] ,\text{ and }\left[
\begin{array}{c}
c_{1,\text{out}} \\
c_{2,\text{out}}%
\end{array}%
\right] =\left[
\begin{array}{cc}
\frac{1}{\sqrt{2}} & \frac{1}{\sqrt{2}} \\
\frac{1}{\sqrt{2}} & -\frac{1}{\sqrt{2}}%
\end{array}%
\right] \left[
\begin{array}{c}
c_{\uparrow ,\text{out}}^{\prime } \\
c_{\downarrow ,\text{out}}^{\prime }%
\end{array}%
\right].  \label{eq:IO1}
\end{eqnarray}%
The Sagnac interferometer is weakly coupled with the cavity, thus we have
\begin{eqnarray}
\left[
\begin{array}{c}
c_{\uparrow ,\text{out}}^{\prime } \\
c_{\downarrow ,\text{out}}^{\prime }%
\end{array}%
\right] =\left[
\begin{array}{c}
e^{i\varphi }\sqrt{1-\epsilon ^{2}}c_{\uparrow ,\text{in}}^{\prime
}+\epsilon c_{\uparrow } \\
e^{i\varphi }\sqrt{1-\epsilon ^{2}}c_{\downarrow ,\text{in}}^{\prime
}+\epsilon c_{\downarrow }%
\end{array}%
\right],  \label{eq:IO2}
\end{eqnarray}%
where $\epsilon $ is the coupling rate and $\varphi $ is the phase delay of
the Sagnac interferometer. We choose the gauge such that the phase for $%
c_{\uparrow }$ and $c_{\downarrow }$ in the above equation is zero.
According to Eqs.~\ref{eq:IO1} and \ref{eq:IO2}, we obtain
\begin{eqnarray}
\left[
\begin{array}{c}
c_{1,\text{out}} \\
c_{2,\text{out}}%
\end{array}%
\right] =\left[
\begin{array}{c}
e^{i\varphi }\sqrt{1-\epsilon ^{2}}c_{1,\text{in}}+\epsilon c_{+} \\
e^{i\varphi }\sqrt{1-\epsilon ^{2}}c_{2,\text{in}}+\epsilon c_{-}%
\end{array}%
\right],
\end{eqnarray}%
with $c_{\pm }=\frac{c_{\uparrow }\pm c_{\downarrow }}{\sqrt{2}}$. We see
that the input-output port 1 and port 2 are coupled with the spin states $%
|\pm \rangle $ respectively, which allow us to excite and measure the
photons in the $\tau ^{x}$ basis. To measure the biorthogonal chiral
displacement, we first prepare the initial state $|\phi_0\rangle$ by
exciting the $|+\rangle$ state at A site of unit cell $n=0$ with a narrow
pulse, then let the system evolve to $|\phi_t\rangle$ at time $t$. The
output field intensities of the two ports at time $t$ are proportional to
the photon field intensities in state $|\pm\rangle$, respectively. We can
measure the output field intensities at each site and obtain $I_1 (n,\alpha)$
for port 1, and $I_2 (n,\alpha)$ for port 2, with $\alpha=0,1$ corresponding
to $A,B$ sublattice sites respectively. Then, Eq. 6 in the main text can be
written as: $\mathcal{C}(t)=\frac{2}{I_\text{tot}} \sum_nn[I_1 (n,0)+I_2
(n,1)-I_2 (n,0)-I_1 (n,1)]$, with total intensity $I_\text{tot}=\sum_nI_1
(n,0)+I_2 (n,0)+I_1 (n,1)+I_2 (n,1)$.

In realistic experiments, fabrication of large arrays of coupled micro-ring
cavities is a mature technology and using auxiliary coupling waveguides with
gain/loss arms to generate asymmetric tunneling has been demonstrated in a
very recent experiment~\cite{science.aba8996}. Moreover, the clean
non-Hermitian SSH Hamiltonian has also been realized in electronic circuits
with site addressability~\cite{arXiv1907.11562}, where disorders can be
introduced simply. With two SSH electronic circuits, we can probe the bulk
topology (based on the Ramsey interferometer scheme) by measuring the
interference between local resonators from the two circuits (i.e., in the $%
\tau^x$ basis), and this can be done since both amplitude and phase of the
local resonators can be measured. Therefore, the experimental realization
and detection of our non-Hermitian topological Anderson states should not
represent a major difficulty.

\begin{figure}[tb]
\includegraphics[width=0.7\linewidth]{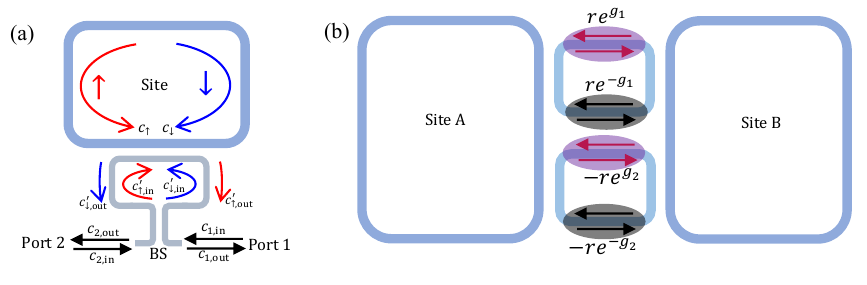}
\caption{(a) Input-output waveguide for measuring photon distributions in
the spin basis $|\pm \rangle $. (b) Site cavities coupled by two coupler
cavities which allow the full tunability of the tunnelings through gain and
loss control.}
\label{figS3:input-output}
\end{figure}

\section{Appendix C: The winding number}
As we defined in the main
text, the winding number is
\begin{eqnarray}
\nu&=&\frac{1}{4}\mathcal{T}\{Q\Gamma[\hat{X},Q]+Q^\dag\Gamma[\hat{X},Q^\dag]%
\}=\frac{1}{4}\mathcal{T}\{Q\Gamma[\hat{X},Q]\}+h.c.,  \label{eq:windingS}
\end{eqnarray}
where the first term $\mathcal{T}\{Q\Gamma[\hat{X},Q]\}$ corresponds to
directly applying the Hermitian formula to the flattened non-Hermitian
Hamiltonian $Q$, this term is not guaranteed to be real for a general
non-Hermitian system since $Q\neq Q^\dag$.
%The definition in which is given by
%the first term may only apply to certain special non-Hermitian systems.
In our model, we find that $\mathcal{T}\{Q\Gamma[\hat{X},Q]\}$ may have a
small imaginary part in the PT-symmetry breaking region for each disorder
configuration, and for parameters away from the phase transition points, its
disorder average can be nearly real with imaginary part $\ll1$.

We have considered the open boundary conditions in this paper, and a naive
evaluation of the trace in calculating $\nu $ yields identical zero, since
the contribution from the bulk states is canceled exactly by the boundary
modes (we will discuss this in detail at the end of this section). Here we
follow the idea in Ref.~\cite%
{PhysRevLett.113.046802,PhysRevB.89.224203,science.aat3406}, and evaluate
the trace per volume in Eq.~\ref{eq:windingS} over the central part of the
lattice chain (Here we exclude 100 lattice sites from each ends of the
chain). Nevertheless, we find that, beside the strong fluctuations at the
phase boundaries, the winding number also fluctuates more strongly in the
PT-symmetry breaking region than the PT-symmetric region. The reasons are
discussed in the following.

For a clean system, it is known that the PT-symmetry breaking is accompanied
by the appearance of exceptional points where more than one right (left)
eigenstates can coalesce and the Hamiltonian is defective. For the disorder
driven PT-symmetry breaking studied here, it is also possible that the
Hamiltonian becomes defective for certain specific disorder configurations
(with fixed disorder strength). A defective Hamiltonian can not be flattened
in the form of Eq.~(1) in the main text. Fortunately, if we consider a
finite lattice with finite disorder configurations, the probability for the
random Hamiltonian to be defective is zero, as confirmed by our numerical
simulations. On the other hand, in the thermodynamic limit (where the system
is infinite), we can always get the specific disorder configuration at
certain spatial interval such that the Hamiltonian is defective. In this
case, we have infinite number of eigenstates (localized) and the probability
for a state to coalesce with others is zero. As a result, we can safely
exclude these defective disorder configurations in the calculation of the
winding number as long as their occurrence probability is zero, which shares
the same spirit as that one can exclude the exceptional points from the
integral over the Brillouin zone when calculating the clean-system winding
number~\cite{PhysRevLett.121.213902}. In actual practice, we do not need to
exclude any disorder configurations in the numerical simulations, as we
adopt a finite lattice and consider finite number of disorder
configurations, where the Hamiltonian is found to be always non-defective.

Though the Hamiltonian is always non-defective in the numerical simulations,
we do find that there are more chances to obtain two nearly coalescing
states (where the Hamiltonian is close to be defective) in the gapless
PT-symmetry breaking region than other regions. Let us denote such two
nearly identical states as $|\Psi _{1}^{R}\rangle $, $|\Psi _{2}^{R}\rangle $
and the corresponding left eigenstates as $|\Psi _{1}^{L}\rangle $, $|\Psi
_{2}^{L}\rangle $, with energy given by $E_{d}$. The biorthonormality
requires that $\langle \Psi _{1}^{L}|\Psi _{2}^{R}\rangle =0$ but $\langle
\Psi _{1}^{L}|\Psi _{1}^{R}\rangle =1$. Recall that $|\Psi _{1}^{R}\rangle $
and $|\Psi _{2}^{R}\rangle $ are nearly identical, therefore, the
distributions $P_{1}(n,\alpha )=\langle \Psi _{1}^{L}(n,\alpha )|\Psi
_{1}^{R}(n,\alpha )\rangle $ must have amplitudes $|P_{1}(n,\alpha )|$ much
larger than $1$ for some sites (the chiral symmetry $\sum_{n}P_{1}(n,0)=%
\sum_{n}P_{1}(n,1)=0.5$ is still satisfied). Similar analysis applies to
state $|\Psi _{2}^{R}\rangle $. These properties of $|\Psi _{1,2}^{R}\rangle
$ would lead to stronger fluctuations in the calculated winding number,
especially when \textit{i}) $P_{1}(n,\alpha )$ is occasionally distributed
around the boundary of the central trace volume, or \textit{ii}) $%
E_{d}\simeq 0$ since $|P_{1}(n,0)|$ would be much larger or smaller than $%
|P_{1}(n,1)|$, which is also the reason why the fluctuation becomes more
significant after the band gap closing at $E=0$, or \textit{iii}) the system
size is too small. Though the fluctuations of the calculated winding numbers
are stronger than the Hermitian models, clear phase boundaries can still be
obtained, as shown in Fig.~2 in the main text. The calculated winding number
is mainly distributed near the averaged value $\nu =1$ ($\nu =0$) in the
topological (trivial) phase, as further confirmed in Fig.~\ref{figS1:band}%
(d) where more disorder configurations ($\sim 1000$) are considered.

The existence of zero edge modes can be determined analytically using $%
\Lambda _{\alpha }$. While for the winding number $\nu $, it is hard to
obtain an analytic formula as a function of system parameters (even for
Hermitian disorders). To connect the existence of zero edge modes and
non-trivial real-space winding, we define the biorthogonal chiral
displacement of the two zero edge modes (one on each boundary), which reads
\begin{equation}
\mathcal{C}_{\text{edge}}=\frac{1}{4N}\sum_{\alpha }\langle \Psi _{0,\alpha
}^{L}|\Gamma \hat{X}|\Psi _{0,\alpha }^{R}\rangle +h.c.
\end{equation}%
with $\alpha =0,1$ the sublattice index and $2N$ the total number of unit
cells. $|\Psi _{0,\alpha }^{L,R}\rangle $ corresponds to the zero edge mode
occupying sublattice site $\alpha $, thus it is located at the boundary end
with sublattice site $\alpha $. If there exist two zero edge modes, we have $%
\sum_{\alpha }\langle \Psi _{0,\alpha }^{L}|\Gamma \hat{X}|\Psi _{0,\alpha
}^{R}\rangle =\langle \Psi _{0,0}^{L}|\hat{X}|\Psi _{0,0}^{R}\rangle
-\langle \Psi _{0,1}^{L}|\hat{X}|\Psi _{0,1}^{R}\rangle \simeq -2N$, and
thereby $\mathcal{C}_{\text{edge}}=-1$. If there is no zero edge mode, we
simply have $\mathcal{C}_{\text{edge}}=0$.

On the other hand, the real-space winding number $\nu $ is encoded in the
localized bulk states. Suppose we can separate the bulk states from the zero
edge modes, so we can use the bulk states (i.e., the flattened Hamiltonian $Q
$ only contains the bulk states) to calculate the real-space winding number
defined in Eq.~\ref{eq:windingS} with the trace per volume evaluated over
the whole lattice chain. Then the winding number equals to the bulk
biorthogonal chiral displacement,
\begin{eqnarray}
\nu  &=&\frac{1}{4}\mathcal{T}\{Q\Gamma \lbrack \hat{X},Q]\}+h.c.=\frac{1}{2}%
\mathcal{T}\{Q\Gamma \hat{X}Q\}+h.c.  \nonumber \\
&=&\frac{1}{4N}\sum_{j,s}\langle \Psi _{j,s}^{L}|\Gamma \hat{X}|\Psi
_{j,s}^{R}\rangle +h.c.,
\end{eqnarray}%
where $j,s$ run over the bulk states in the summation, and we have used $%
Q\Gamma =-\Gamma Q$ in the derivation. We find that $\mathcal{C}_{\text{edge}%
}+\nu =0$ if the Hamiltonian is non-defective because $\mathcal{C}_{\text{%
edge}}+\nu =\frac{1}{2}\mathcal{T}\{\mathbb{I}\Gamma \hat{X}\}+h.c.$ and $%
\mathbb{I}=\sum_{j,s}|\Psi _{j,s}^{R}\rangle \langle \Psi
_{j,s}^{L}|+\sum_{\alpha }|\Psi _{0,\alpha }^{R}\rangle \langle \Psi
_{0,\alpha }^{L}|$ satisfies $\mathbb{I}|\phi \rangle =|\phi \rangle $ for
arbitrary $|\phi \rangle $ if the Hamiltonian is non-defective. Fortunately,
if we consider a finite lattice with finite disorder configurations, the
probability for the random Hamiltonian to be defective is zero, as confirmed
by our numerical simulations. Therefore, the existence of zero edge modes
(i.e., $\mathcal{C}_{\text{edge}}=-1$) is connected to non-trivial
real-space winding number $\nu =1$ encoded in the bulk states. We have
numerically verified $\mathcal{C}_{\text{edge}}$ and bulk $\nu $ in the
gapped region where the bulk states can be separated from zero edge modes.

We want to point out that, the non-Hermitian topology is encoded in the bulk
states under open boundary condition supporting zero edge states.
Numerically, it is hard to separate the zero edge states from bulk states in
the strong disorder region with gapless bulk spectrum. Instead, we keep all
the states and directly evaluate the trace per volume in Eq.~\ref%
{eq:windingS} over the central part of the system to eliminate the effects
of zero edge modes.

\begin{figure}[tb]
\includegraphics[width=1.0\linewidth]{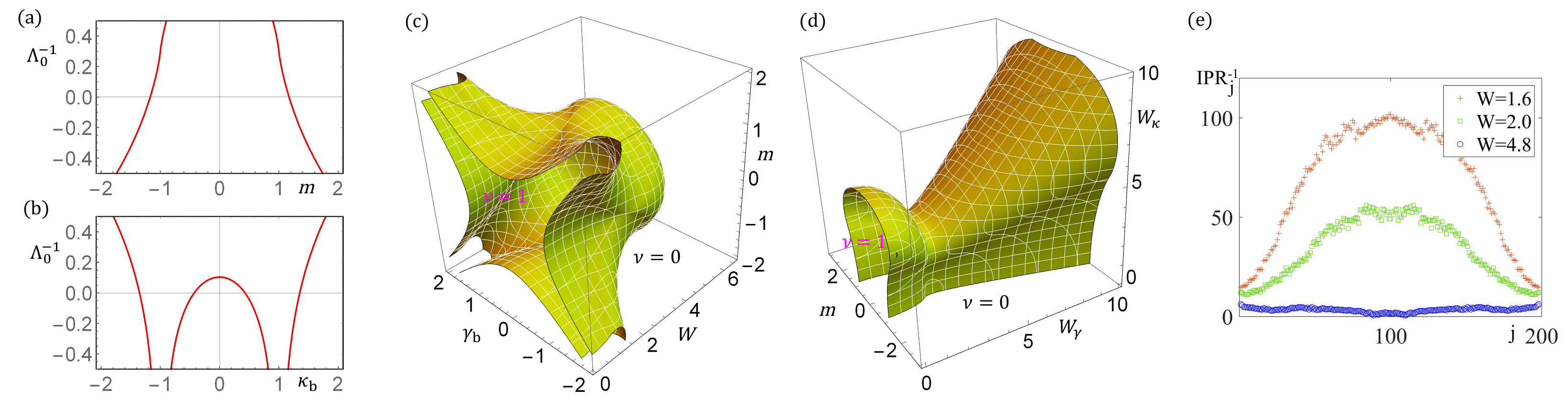}
\caption{(a) (b) Zero-energy-mode biorthogonal localization length as
functions of $m$ (with $\protect\kappa _{\text{b}}=0.1$) and $\protect\kappa %
_{\text{b}}$ (with $m=1.1$). $W=W_{\protect\gamma }=8W_{\protect\kappa }=2$.
(c) (d) The delocalized critical surfaces on the parameter space ($m$, $%
\protect\gamma _{\text{b}}$, $W$) and ($m$, $W_{\protect\kappa }$, $W_{%
\protect\gamma }$) with $W=W_{\protect\gamma }=8W_{\protect\kappa }$. (e)
The IPR of all bulk eigenstates at different disorder strengths $W=W_{%
\protect\gamma }=8W_{\protect\kappa }$. $j$ is the eigenvalue index and the
system contains 200 lattice sites ($N=50$). Other parameters are the same as Fig.~3c in the main text.}
\label{figS2:phase}
\end{figure}

\section{Appendix D: Biorthogonal localization length}
Different from
Hermitian systems, here the eigenstates (both left and right ones) cannot be
interpreted as the probability amplitude of particle's
%\U{62af}
distribution.
In certain parameter regions (including certain phase transition points), we
may even have all left/right eigenstates localized at the boundary due to
non-Hermitian skin effects. It is a priori unclear how to characterize the
localization-delocalization criticality for the non-Hermitian disorder
systems. Given the fact that the left and right eigenstates are
inequivalent, we need to take into account both of them to characterize the
localization. For the zero-energy modes, the eigenequations $H|\Psi
_{0}^{R}\rangle =0$ and $H^{\dag }|\Psi _{0}^{L}\rangle =0$ are
\begin{eqnarray}
0 &=&(J-\kappa _{n})|\Psi _{0}^{R}(n+1,0)\rangle +(m+\gamma _{n})|\Psi
_{0}^{R}(n,0)\rangle   \nonumber \\
0 &=&(m-\gamma _{n+1})|\Psi _{0}^{R}(n+1,1)\rangle +(J+\kappa _{n})|\Psi
_{0}^{R}(n,1)\rangle   \nonumber \\
0 &=&(J+\kappa _{n})|\Psi _{0}^{L}(n+1,0)\rangle +(m-\gamma _{n})|\Psi
_{0}^{L}(n,0)\rangle   \nonumber \\
0 &=&(m+\gamma _{n+1})|\Psi _{0}^{L}(n+1,1)\rangle +(J-\kappa _{n})|\Psi
_{0}^{L}(n,1)\rangle .  \label{eq:zero_modeS}
\end{eqnarray}%
The solutions are
\begin{eqnarray}
|\Psi _{0}^{R}(n+1,\alpha )\rangle  &=&-\left[ \frac{m+(-1)^{\alpha }\gamma
_{n+\alpha }}{J-(-1)^{\alpha }\kappa _{n}}\right] ^{\eta _{\alpha }}|\Psi
_{0}^{R}(n,\alpha )\rangle   \nonumber \\
|\Psi _{0}^{L}(n+1,\alpha )\rangle  &=&-\left[ \frac{m-(-1)^{\alpha }\gamma
_{n+\alpha }}{J+(-1)^{\alpha }\kappa _{n}}\right] ^{\eta _{\alpha }}|\Psi
_{0}^{L}(n,\alpha )\rangle ,  \label{eq:zero_modeS2}
\end{eqnarray}%
with $\alpha =0,1$ and $\eta _{\alpha }=\pm 1$ corresponding to sublattice
sites $A$ and $B$, respectively. We obtain the biorthogonal distributions $%
P(n,\alpha )=\langle \Psi _{0}^{L}(n,\alpha )|\Psi _{0}^{R}(n,\alpha
)\rangle $ as given in the main text (by setting $J=1$ as the energy unit),
\begin{equation}
P(n,\alpha )=P(0,\alpha )\prod_{n^{\prime }=0}^{n-1}\left( \frac{m-\gamma
_{n^{\prime }+\alpha }^{2}}{1-\kappa _{n^{\prime }}^{2}}\right) ^{\eta
_{\alpha }}.
\end{equation}%
The zero-energy-mode biorthogonal localization lengths are
\begin{eqnarray}
\Lambda _{\alpha }^{-1} &=&-\frac{1}{2}\lim_{n\rightarrow \infty }\frac{1}{n}%
\ln \left\vert \frac{P(n,\alpha )}{P(0,\alpha )}\right\vert   \nonumber \\
&=&\lim_{n\rightarrow \infty }\frac{\eta _{\alpha }}{2n}\sum_{n^{\prime
}=0}^{n-1}\left( \ln |1-\kappa _{n^{\prime }}^{2}|-\ln |m-\gamma _{n^{\prime
}+\alpha }^{2}|\right) .
\end{eqnarray}

According to Birkhoff's ergodic theorem, we can use the ensemble average to
evaluate $\Lambda _{\alpha }$
\begin{eqnarray}
\Lambda _{0}^{-1}=-\Lambda _{1}^{-1} &=&\frac{1}{2}\int_{\gamma _{\text{b}%
}-W_{\gamma }/2}^{\gamma _{\text{b}}+W_{\gamma }/2}d\gamma \int_{\kappa _{%
\text{b}}-W_{\kappa }/2}^{\kappa _{\text{b}}+W_{\kappa }/2}d\kappa \left(
\ln |1-\kappa ^{2}|-\ln |m-\gamma ^{2}|\right)  \nonumber \\
&=&\sum_{s,s^{\prime }=\pm }\left[ \ln \frac{|W_{\kappa }+2s+2ss^{\prime
}\kappa _{\text{b}}|^{\frac{s+ss^{\prime }\kappa _{\text{b}}}{2W_{\kappa }}%
}|(W_{\kappa }+2s\kappa _{\text{b}})^{2}-4|^{1/8}}{|W_{\gamma
}+2ms+2ss^{\prime }\gamma _{\text{b}}|^{\frac{ms+ss^{\prime }\gamma _{\text{b%
}}}{2W_{\gamma }}}|(W_{\gamma }+2s\gamma _{\text{b}})^{2}-4m^{2}|^{1/8}}%
\right] .
\end{eqnarray}%
Follow a similar analysis as in Ref.~\cite%
{PhysRevLett.113.046802,PhysRevB.89.224203,science.aat3406}, we find
that, except for some special cases $|W_{\kappa }\pm 2\kappa _{\text{b}}|=2$
and $|W_{\gamma }\pm 2\gamma _{\text{b}}|=2|m|$, $\Lambda _{\alpha }^{-1}$
is analytic at the critical phase boundary $\kappa _{\text{b}}^{c}$ and $%
m^{c}$ with expansion $\Lambda _{0}^{-1}(m,\kappa _{\text{b}%
})=a_{m}(m-m^{c})+a_{\kappa }(\kappa _{\text{b}}-\kappa _{\text{b}%
}^{c})+\cdots $ [as shown in Figs.~\ref{figS2:phase}(a) and \ref{figS2:phase}%
(b)], leading to the critical exponent $1$. For the special critical cases,
one has $\Lambda _{0}^{-1}(m)\sim (m-m^{c})\ln |m-m^{c}|$ with fixed $\kappa
_{\text{b}}$, or $\Lambda _{0}^{-1}(\kappa _{\text{b}})\sim (\kappa _{\text{b%
}}-\kappa _{\text{b}}^{c})\ln |\kappa _{\text{b}}-\kappa _{\text{b}}^{c}|$
with fixed $m$.

In Figs.~\ref{figS2:phase}(c) and \ref{figS2:phase}(d), we also plot the
phase diagrams in the parameter space ($m$, $\gamma _{\text{b}}$, $W$) and ($%
m$, $W_{\kappa }$, $W_{\gamma }$) obtained from $\Lambda _{0}$. Here $%
W=W_{\gamma }=8W_{\kappa }$. We see that the system is topological (trivial)
in the $W_{\kappa }\rightarrow \infty $ ($W_{\gamma }\rightarrow \infty $).
Due to the competition between Hermitian ($m$) and non-Hermitian ($\gamma _{%
\text{b}}$) tunnelings which tend to weaken each other, the system stays in
the topological phase up to very large $m$ and $\gamma _{\text{b}}$ along
the directions $|m|=|\gamma _{\text{b}}|$.

When $E\neq 0$, it is difficult to obtain the analytic expression of the
localization length. Here we numerically calculate the biorthogonal
localization lengths using the transfer matrix method~\cite%
{RevModPhys.80.1355}, where the transfer matrix for right (left)
eigenstates can be obtained using $H$ ($H^{\dag }$). The eigenequations $%
H|\Psi _{E}^{R}\rangle =E|\Psi _{E}^{R}\rangle $ and $H^{\dag }|\Psi
_{E}^{L}\rangle =E^{\ast }|\Psi _{E}^{L}\rangle $ can be written as
\[
\left[
\begin{array}{c}
\Psi _{E}^{R}(\widetilde{n},0) \\
\Psi _{E}^{R}(\widetilde{n},1)%
\end{array}%
\right] =M_{\widetilde{n}}^{R}\left[
\begin{array}{c}
\Psi _{E}^{R}(0,0) \\
\Psi _{E}^{R}(0,1)%
\end{array}%
\right] ,\text{ with }M_{\widetilde{n}}^{R}=\prod_{n=0}^{\widetilde{n}-1}%
\left[
\begin{array}{cc}
-\frac{m+\gamma _{n}}{1-\kappa _{n}} & \frac{E}{1-\kappa _{n}} \\
-\frac{E(m+\gamma _{n})}{(1-\kappa _{n})(m-\gamma _{n+1})} & \frac{%
E^{2}-1+\kappa _{n}^{2}}{(1-\kappa _{n})(m-\gamma _{n+1})}%
\end{array}%
\right]
\]%
and similarly for $|\Psi _{E}^{L}\rangle $ with
\[
M_{\widetilde{n}}^{L}=\prod_{n=0}^{\widetilde{n}-1}\left[
\begin{array}{cc}
-\frac{m-\gamma _{n}}{1+\kappa _{n}} & \frac{E^{\ast }}{1+\kappa _{n}} \\
-\frac{E^{\ast }(m-\gamma _{n})}{(1+\kappa _{n})(m+\gamma _{n+1})} & \frac{%
E^{\ast 2}-1+\kappa _{n}^{2}}{(1+\kappa _{n})(m+\gamma _{n+1})}%
\end{array}%
\right] .
\]%
The biorthogonal localization length is defined as
\begin{equation}
\Lambda _{E}^{-1}=\lim_{\widetilde{n}\rightarrow \infty }\frac{\ln |\lambda
_{\widetilde{n}}^{R}\lambda _{\widetilde{n}}^{L}|}{2\widetilde{n}},
\end{equation}%
with $\lambda _{\widetilde{n}}^{R}$ ($\lambda _{\widetilde{n}}^{L}$) the
larger eigenvalue of $M_{\widetilde{n}}^{R}$ ($M_{\widetilde{n}}^{L}$)~\cite%
{Z.Phys.B.53,PhysRevB.29.3111}. Our numerical results are shown in
Fig.~3(c) in the main text, which indicate that all states with $E\neq 0$
are localized in every instance with disorders. At $E=0$, $\Lambda _{E}^{-1}$
is consistent with the analytic zero-energy-mode biorthogonal localization
length.

The localization properties can also be characterized by the inverse
participation ratio (IPR) of the eigenstates, and larger IPR corresponds to
stronger localization~\cite{RevModPhys.80.1355}. Here we define the
biorthogonal IPR as
\begin{equation}
\text{IPR}_{\widetilde{j}}=\frac{\sum_{n,\alpha}|\langle\Psi^L_{\widetilde{j}%
}(n,\alpha)|\Psi^R_{\widetilde{j}}(n,\alpha)\rangle|^4}{\sum_{n,\alpha}|%
\langle\Psi^L_{\widetilde{j}}(n,\alpha)|\Psi^R_{\widetilde{j}%
}(n,\alpha)\rangle|^2},
\end{equation}
with $\widetilde{j}=(j,\pm)$ the eigenvalue index. The averaged IPR is
defined as $\text{IPR}=\frac{1}{L}\sum_{\widetilde{j}}\text{IPR}_{\widetilde{%
j}}$ with $L$ the total number of sites. In Fig.~\ref{figS2:phase}(e), we
plot the IPR$_{\widetilde{j}}$ near the phase boundaries for all bulk states
with $E\neq0$, which are well above $1/L$, indicating strong localization.

\section{Appendix E: Biorthogonal chiral displacement}
In this
section, we prove that the averaged Biorthogonal chiral displacement
converges to the winding number in the PT-symmetric region. In the presence
of disorder, the trace over the whole system may be replaced by a disorder
average over a single unit cell according to Birkhoff's ergodic theorem~\cite%
{PhysRevLett.113.046802,PhysRevB.89.224203,science.aat3406}. We can
evaluate the winding number at unit cell $n=0$, which is
\begin{eqnarray}
\nu (0) &=&\frac{1}{4}\sum_{\alpha =A,B}\langle 0_{\alpha }|Q\Gamma \lbrack
\hat{X},Q]|0_{\alpha }\rangle +h.c.  \nonumber \\
&=&\frac{1}{4}\sum_{\alpha }\langle 0_{\alpha }|Q\Gamma \hat{X}Q|0_{\alpha
}\rangle +h.c.  \nonumber \\
&=&\frac{1}{4}\sum_{\alpha }\langle 0_{\alpha }|(\mathbb{I}-2P_{-})\Gamma
\hat{X}(\mathbb{I}-2P_{-})|0_{\alpha }\rangle +h.c.,  \label{eq:windingS0}
\end{eqnarray}%
with $\mathbb{I}=P_{+}+P_{-}$. In the PT-symmetric region (without
eigenstates coalescing), the eigenstates form a complete biorthonormal
basis, and an arbitrary state $|\phi \rangle $ can be expanded as $|\phi
\rangle =\sum_{s=\pm ,j}\phi _{j,s}|\Psi _{j,s}^{R}\rangle $ with $\phi
_{j,s}=\langle \Psi _{j,s}^{L}|\phi \rangle $, and $\mathbb{I}|\phi \rangle
=|\phi \rangle $. Therefore, we have
\begin{eqnarray}
\nu (0) &=&\sum_{\alpha }\langle 0_\alpha |P_{-}\Gamma \hat{X}P_{-}|0_\alpha
\rangle +h.c.  \nonumber \\
&=&\sum_{\alpha }\left[ \frac{1}{2}\sum_{s=\pm ,j}a_{j,s}^{R}(\alpha
)a_{j,s}^{L}(\alpha )\langle \Psi _{j,s}^{L}|\Gamma \hat{X}|\Psi
_{j,s}^{R}\rangle +\sum_{j\neq j^{\prime }}a_{j,-}^{R}(\alpha )a_{j^{\prime
},-}^{L}(\alpha )\langle \Psi _{j,-}^{L}|\Gamma \hat{X}|\Psi _{j^{\prime
},-}^{R}\rangle \right] +h.c.,
\end{eqnarray}%
with $a_{j,s}^{R}(\alpha )=\langle 0_{\alpha }|\Psi _{j,s}^{R}\rangle $ and $%
a_{j,s}^{L}(\alpha )=\langle \Psi _{j,s}^{L}|0_{\alpha }\rangle $. Similar
to the Hermitian system, we find (numerically) that the off-diagonal part
(the sum over $j\neq j^{\prime }$) in the above equation provides very small
contributions, and
\[
\nu (0)\simeq \frac{1}{2}\sum_{\alpha ,s=\pm ,j}a_{j,s}^{R}(\alpha
)a_{j,s}^{L}(\alpha )\langle \Psi _{j,s}^{L}|\Gamma \hat{X}|\Psi
_{j,s}^{R}\rangle +h.c.
\]

The biorthogonal chiral displacement is defined as
\begin{eqnarray}
&&\mathcal{C}_A(t)=\langle L(t)|\Gamma\hat{X}|R(t)\rangle +h.c. = \langle
0_A|\exp(iHt)\Gamma\hat{X}\exp(-iHt)|0_A\rangle +h.c.  \nonumber \\
&&=\sum_{j,s}a^R_{j,s}(A)a^L_{j,s}(A)\langle\Psi_{j,s}^L|\Gamma\hat{X}%
|\Psi_{j,s}^R\rangle +\sum_{(j,s)\neq(j^{\prime },s^{\prime
})}a^R_{j,s}(A)a^L_{j^{\prime },s^{\prime }}(A)e^{i(E_{j,s}-E_{j^{\prime
},s^{\prime }})t}\langle\Psi_{j,s}^L|\Gamma\hat{X}|\Psi_{j^{\prime
},s^{\prime }}^R\rangle +h.c.  \label{Eq:ChiralS}
\end{eqnarray}
In the PT-symmetric region, all eigenvalues $E_{j,s}$ are real. Therefore,
the second term of Eq.~\ref{Eq:ChiralS} rapidly oscillates, which converges
to zero when averaged over sufficiently long time sequences. The first term
of Eq.~\ref{Eq:ChiralS} is time independent and gives the mean chiral
displacement
\begin{eqnarray}
\bar{\mathcal{C}}_A(\infty)&=&\sum_{j,s}a^R_{j,s}(A)a^L_{j,s}(A)\langle%
\Psi_{j,s}^L|\Gamma\hat{X}|\Psi_{j,s}^R\rangle+h.c.
\end{eqnarray}
Similarly, we can calculate the chiral displacement $\bar{\mathcal{C}}%
_B(\infty)$ for initial state $|0_B\rangle$ and $\nu(0)=\bar{\mathcal{C}}%
(\infty)\equiv\frac{1}{2}[\bar{\mathcal{C}}_A(\infty)+\bar{\mathcal{C}}%
_B(\infty)]$. For Hermitian systems, one has $\bar{\mathcal{C}}_A(\infty)=%
\bar{\mathcal{C}}_B(\infty)$~\cite{science.aat3406}, and we find it also
holds here. We can just use the chiral displacement $\mathcal{C}_A(t)$ to
probe the topology and drop the subscript $A$.

\section{Appendix F: Effects of interaction}
Generally, interaction is
irrelevant for the photonic systems since photons do not interact with each
other. While for atomic system, strong interaction is possible. If the bulk
states have a gap around $E=0$, we expect that the topology and zero edge
modes are not affected by interactions that are small compared to the gap.
For the disordered system, the gap might be an averaged result (i.e., the
probability to find bulk states vanishes within the gap around). On the
other hand, the topology may be inhibited by interaction if the interaction
is too strong or if the gap at $E=0$ disappears (which is typical for the
topological states in strong disorder region). Further numerical simulations
are needed to ascertain the interacting physics. One possible way is to
calculate the topological invariant based on the many-body ground state (as
defined in the following). Numerically, the many-body ground state can be
obtained by exact diagonalization method, which may not be practicable since
a large system is required to suppress the boundary effects. It is also
possible to generalize the density matrix renormalization group method to
non-Hermitian systems. We believe that the interaction effects in such
non-Hermitian topological Anderson insulators are more subtle than that in
the ordinary topological insulators, which is a very interesting direction
worth to be addressed in future works.

Now we give a possible generalization of the winding number. We note that
the non-interacting winding number equals to the biorthogonal chiral
displacement averaged over occupied bulk states (i.e., $\nu =\frac{1}{2N}%
\sum_{j}\langle \Psi _{j,-}^{L}|\Gamma \hat{X}|\Psi _{j,-}^{R}\rangle +h.c.$
due to $\langle \Psi _{j,+}^{L}|\Gamma \hat{X}|\Psi _{j,+}^{R}\rangle
=\langle \Psi _{j,-}^{L}|\Gamma \hat{X}|\Psi _{j,-}^{R}\rangle $).
Therefore, we could define a many-body winding number as
\begin{equation}
\nu _{\text{int}}=\frac{1}{2N}\langle G^{L}|\Gamma X|G^{R}\rangle +h.c.,
\end{equation}%
with $|G^{L,R}\rangle $ the left/right many-body ground state. The ground
state has lowest real energy for a given particle number (which is still a
good quantum number). If the lowest real level contains two or more states,
we choose the one with lowest imaginary energy among them. For
non-interacting particles, bosons and fermions share the same
single-particle spectrum, zero edge modes and topological invariant. While
for interacting many-body ground state, we need discuss bosons and fermions
separately. It can be shown that, for fermions, the above many-body
invariant with $2N-1$ interacting particles can be reduced to the
non-interacting invariant if the non-interacting bulk states have a gap at $%
E=0$ (this is because, if we have $E=0$ bulk states, the particles in $%
|G^{L,R}\rangle $ may occupy zero edge modes instead of the zero bulk
states). For the disordered system, this generalization is consistent with
non-interacting invariant if the probability to find zero bulk states
vanishes for a finite size lattice and a finite number of disorder
configurations. While for bosons, if the interaction is weak, $%
|G^{L,R}\rangle $ may corresponds to occupation of all particles on a few
low single-particle levels, which should be trivial. Interesting bosonic
many-body topological states may emerge under strong interactions, for
example, the many-body invariant of simple hardcore bosons (which can be
mapped to non-interacting fermions that preserves density) is reduced to
non-interacting fermion invariant.

In the following, we show how to reduce the many-body invariant in the
non-interaction limit. The Hamiltonian can be written as
\[
H=\sum_{j,s}E_{j,s}\beta _{R,j,s}^{\dag }\beta _{L,j,s},
\]%
where $\beta _{L,j,s}=\sum_{n,\alpha }\mathcal{L}_{j,s;n,\alpha }c_{n,\alpha
}$ and $\beta _{R,j,s}^{\dag }=\sum_{n,\alpha }c_{n,\alpha }^{\dag }\mathcal{%
R}_{n,\alpha ;j,s}$, with $\mathcal{L}_{j,s;n,\alpha }=\langle \Psi
_{j,s}^{L}|n,\alpha \rangle $ and $\mathcal{R}_{n,\alpha ;j,s}=\langle
n,\alpha |\Psi _{j,s}^{R}\rangle $. They satisfy $\beta _{L,j,s}\beta
_{R,j^{\prime },s^{\prime }}^{\dag }\pm \beta _{R,j^{\prime },s^{\prime
}}^{\dag }\beta _{L,j,s}=\delta _{j,j^{\prime }}\delta _{s,s^{\prime }}$,
with $\pm $ corresponding to fermions and bosons, respectively.

\emph{Fermions:} The
non-interacting fermion ground state with $N_{p}$ (we assume $N_{p}\leq 2N$)
particles is
\begin{equation}
|G^{R}\rangle =\prod_{j=1}^{N_{p}}\beta _{R,j,-}^{\dag }|\text{vac}\rangle
\text{ and }\langle G^{L}|=\langle \text{vac}|\prod_{j=1}^{N_{p}}\beta
_{L,j,-}
\end{equation}%
with ground state energy $\sum_{j}^{N_{p}}E_{j,-}$, here $|\text{vac}\rangle
$ the vacuum state. If the Hamiltonian is non-defective, we have $\mathcal{R}%
_{n,\alpha ;j,s}\beta _{L,j,s}=c_{n,\alpha }$ and $\beta _{R,j,s}^{\dag }%
\mathcal{L}_{j,s;n,\alpha }=c_{n,\alpha }^{\dag }$, thus the many-body
invariant becomes
\begin{eqnarray}
\nu _{\text{int}} &=&\frac{1}{2N}\langle G^{L}|\Gamma X|G^{R}\rangle +h.c.
\nonumber \\
&=&\frac{1}{2N}\langle G^{L}|\sum_{n,\alpha }n(-1)^{\alpha }c_{n,\alpha
}^{\dag }c_{n,\alpha }|G^{R}\rangle +h.c.  \nonumber \\
&=&\frac{1}{2N}\sum_{n,\alpha ,j,j^{\prime },s,s^{\prime }}n(-1)^{\alpha }%
\mathcal{L}_{j,s;n,\alpha }\mathcal{R}_{n,\alpha ;j^{\prime },s^{\prime
}}\langle G^{L}|\beta _{R,j,s}^{\dag }\beta _{L,j^{\prime },s^{\prime
}}|G^{R}\rangle +h.c.  \nonumber \\
&=&\frac{1}{2N}\sum_{j}^{N_{p}}\sum_{n,\alpha }n(-1)^{\alpha }\mathcal{L}%
_{j,-;n,\alpha }\mathcal{R}_{n,\alpha ;j,-}+h.c.  \nonumber \\
&=&\frac{1}{2N}\sum_{j}^{N_{p}}\langle \Psi _{j,-}^{L}|\Gamma \hat{X}|\Psi
_{j,-}^{R}\rangle +h.c.  \nonumber \\
&=&\frac{1}{4N}\sum_{j}^{N_{p}}\sum_{s=\pm }\langle \Psi _{j,s}^{L}|\Gamma
\hat{X}|\Psi _{j,s}^{R}\rangle +h.c.
\end{eqnarray}%
We have used $\langle \Psi _{j,+}^{L}|\Gamma \hat{X}|\Psi _{j,+}^{R}\rangle
=\langle \Psi _{j,-}^{L}|\Gamma \hat{X}|\Psi _{j,-}^{R}\rangle $. If $%
N_{p}=2N-1$ and there is no zero bulk states, then $\nu _{\text{int}}$ is
just the biorthogonal chiral displacement averaged over all single-particle
bulk states, and thus $\nu _{\text{int}}=\nu $.

\emph{Bosons:}The non-interacting boson
ground state with $N_{p}$ particles is
\begin{equation}
|G^{R}\rangle =\frac{1}{\sqrt{N_{p}!}}(\beta _{R,1,-}^{\dag })^{N_{p}}|\text{%
vac}\rangle \text{ and }\langle G^{L}|=\langle \text{vac}|\frac{1}{\sqrt{%
N_{p}!}}(\beta _{L,1,-})^{N_{p}}
\end{equation}%
and the many-body invariant becomes
\begin{equation}
\nu _{\text{int}}\ =\frac{N_{p}}{2N}\langle \Psi _{1,-}^{L}|\Gamma \hat{X}%
|\Psi _{1,-}^{R}\rangle +h.c.,
\end{equation}%
which is not quantized in general.

\end{widetext}

%\begin{acknowledgments}
\noindent \textbf{Acknowledgement.} This work is supported by AFOSR
(FA9550-16-1-0387, FA9550-20-1-0220), NSF (PHY-1806227), ARO
(W911NF-17-1-0128), UTD internal seed grant. X.L. also acknowledges support from the USTC start-up funding.
%\end{acknowledgments}

\noindent \textbf{Data Availability.} The data that support the plots within
this paper and other findings of this paper are available from the
corresponding author upon reasonable request

%\input{Non-Hermitian-TAI_ref.bbl}

%\bibliographystyle{apsrev}
%\bibliographystyle{unsrt}
%\bibliography{Weyl-Paper}
%\newpage
%\newpage

\end{document}